

\tolerance=10000
\input phyzzx

%


\REF\Halp{M.B.Halpern, Phys.Rev.{\bf D19} (1979) 517.}
\REF\XB{
    X. Bekaert and  S. Cucu, hep-th/0104048.}
\REF\Hen{
   X. Bekaert, M. Henneaux and A.   Sevrin, hep-th/0004049.}
\REF\WitUSC{E. Witten, in Proceedings of Strings 95,
hep-th/9507121.}
\REF\Verl{E. Verlinde,  Nucl.Phys. {\bf B455} (1995) 211; hep-th/9506011.}
\REF\cst{C.M. Hull, Nucl.Phys. {\bf B583} (2000) 237, 
hep-th/0004195.}
\REF\csto{C.M. Hull, JHEP {\bf 12} (2000) 007,
hep-th/0011215.}
\REF\gor{M. Goroff and J.H. Schwarz, Phys. Lett. {\bf B127} (1983) 61.} 
\REF\curt{ T. Curtright, Phys. Lett. {\bf 165B} (1985) 304.}
\REF\CF{  T.L. Curtright and P.G.O. Freund, Nucl. Phys. {\bf B172}
(1980) 413.}
\REF\AKO{  C.S. Aulakh, I.G. Koh and S. Ouvry, Phys. Lett. {\bf B173}
(1986) 284.}
\REF\Morr{ J.M.F. Labastida and T.R. Morris, Phys. Lett. {\bf
B180} (1986) 101.}
\REF\KO{ I.G. Koh and S. Ouvry, Phys. Lett. {\bf B179} (1986) 115; J.A.
Garc\'{\i}a and B. Knaepen, Phys. Lett. {\bf B441} (1998) 198.}
\REF\notivarg{
S.~Deser, P.~K.~Townsend and W.~Siegel,
Nucl.\ Phys.\  {\bf B184}, 333 (1981).}
\REF\Sieg{
W. Siegel and  B. Zwiebach, Nucl.Phys.{\bf B282} (1987) 125;
 W. Siegel,  hep-th/0107094.}
\REF\pri{
W. Siegel, private communication.}
\REF\CJP{E. Cremmer, B. Julia, H. Lu and C.N. Pope, Nucl. Phys. {\bf B535 }
(1998)
242, hep-th/9806106.}
\REF\zee{A. Zee, Phys. Rev.
Lett. {\bf 55} (1985) 2379.}
\REF\Nepo{S. Deser and R.I. Nepomechie, Phys. Lett.
{\bf 97A} (1983) 329.}
\REF\Mex{
 H. Garcia-Compean, O. Obregon, J.F.Plebanski, C. Ramirez
       Phys.Rev. D57 (1998) 7501, hep-th/9711115; H. Garcia-Compean, O.
Obregon, C. Ramirez Phys.Rev. D58 (1998) 104012, hep-th/9812175; J.A.
Nieto, Phys. Lett. {\bf A262} (1999) 274, hep-th/9910049.}

%
\font\mybb=msbm10 at 12pt
\def\bbcc#1{\hbox{\mybb#1}}
\def\Z {\bbcc{Z}}
\def\R {\bbcc{R}}

\def \aa {\alpha}
\def \bb {\beta}
\def \gg {\gamma}
\def \dd {\delta}
\def \ee {\epsilon}

\def \kk {\kappa}
\def \ll {\lambda}
\def \mm {\mu}
\def \nn {\nu}

\def \rr {\rho}
\def \ss {\sigma}
\def \tt {\tau}

\def \th {\theta}

\def \lll {\Lambda}

\def \ti {\tilde}

\def \2 {{1 \over 2}}
\def \3 {{1 \over 3}}
\def \4 {{1 \over 4}}
\def \5 {{1 \over 5}}
\def \6 {{1 \over 6}}
\def \7 {{1 \over 7}}
\def \8 {{1 \over 8}}
\def \9 {{1 \over 9}}
\def \00 { \infty}

\def\++ {{(+)}}
\def \- {{(-)}}
\def\+-{{(\pm)}}

\def\ek {\eqn\abc$$}

\def \pa {\partial}
\def\na {\nabla}


 \def\unit{\hbox to 3.3pt{\hskip1.3pt \vrule height 7pt width .4pt
\hskip.7pt
\vrule height 7.85pt width .4pt \kern-2.4pt
\hrulefill \kern-3pt
\raise 4pt\hbox{\char'40}}}

\def\nup#1({Nucl.\ Phys.\  {\bf B#1}\ (}

\def \qq {\qquad}


\Pubnum{ \vbox{ \hbox {QMUL-PH-01-01} \hbox{hep-th/0107149}} }
\pubtype{}
\date{July, 2001}

\titlepage

\title {\bf Duality in  Gravity and Higher Spin Gauge Fields }

\author{C.M. Hull}
\address{Physics Department,
\break
Queen Mary, University of London,
\break
Mile End Road, London E1 4NS, U.K.}
\vskip 0.5cm

\abstract {Dual field theory realisations  are given for linearised gravity in terms of
gauge fields in exotic representations of the Lorentz group. The field equations
and dual representations are discussed for a wide class of higher spin gauge fields.
For non-linear Einstein gravity, such transformations can be implemented locally in
light-cone gauge,
or partially implemented in the presence of a Killing vector.
Sources and the relation to Kaluza-Klein monopoles are discussed.}

\endpage

\chapter{Introduction}

In $D$ dimensions, the  theory of a free abelian vector potential $A_\mm$
can be reformulated as a theory of a free $n$-form gauge field
$\ti A_{\mm_1 \dots \mm_n}$ where $n=D-3$. This gives two dual
formulations of the same theory. In $D=4$, both $A$ and $\ti A$ are
1-forms and the existence of two dual formulations leads to a symmetry of
the equations of motion in which the field strength $F =dA$  transforms  
transforms into its Hodge dual, $*F$.
This becomes part of a $SL(2,\R)$ symmetry of the equations of motion
which acts on both fields and coupling constants, and for certain
supersymmetric theories (such as those with $N=4$ supersymmetry) this is
broken to a discrete $SL(2,\Z)$ symmetry of the quantum theory.

These dualities extend to interacting theories in which $A$ couples to
other fields only through its field strength $F$.
However, in the generalisation to non-abelian Yang-Mills theory, 
or to the minimal coupling to charged matter,
the field equations and Bianchi identities involve the vector 
potential $A$ explicitly, and there seems to be no local covariant  way of 
implementing these duality transformations in classical field theory. In
particular the
$D=4$  electromagnetic duality symmetries appear to be lost -- the $N=4$
super-Yang-Mills classical field equations
only have an $SL(2)$ symmetry in the abelian case.\footnote*{In 
[\Halp], a non-local construction of an electromagnetic dual of Yang-Mills
theory was given, and in [\XB] it was shown that certain duality-invariant
free actions do not admit local covariant non-abelian interactions.} 
Remarkably, there is considerable evidence that in $N=4$ supersymmetric
non-abelian Yang-Mills theories in four dimensions
there is indeed such an $SL(2,\Z)$ S-duality symmetry 
of the full quantum theory 
even though there is no such symmetry of the classical non-abelian 
field equations. This duality  symmetry has had profound implications for 
our understanding of the non-perturbative structure of these
theories.

This S-duality symmetry has a geometric origin.
In 6 dimensions, there  are
interacting (2,0) supersymmetric theories that reduce to
theories containing super-Yang-Mills  in $D<6$.
In the abelian case, the (2,0) theory is a theory of a 2-form gauge field
with self-dual field strength, but the interacting theory cannot be a
local covariant theory of interactions of such fields [\Hen].
Reducing on a 2-torus 
gives a $D=4$ theory with an $SL(2,\Z)$ S-duality symmetry
arising from the diffeomorphisms of the 2-torus [\WitUSC,\Verl]. This $D=4$
theory reduces to super-Yang-Mills in a certain limit, but also inherits
some of the features of the $D=6$ theory that cannot be formulated 
in terms of a local covariant classical field theory.

In these examples, interacting classical field 
theory does not seem to provide a complete description of the full theory
and in particular does not have the appropriate duality symmetries, and
the study of the  free limit and its symmetries
turns out to be a surprisingly good guide to the properties of the full
theory. 

The purpose of this paper is to 
generalise such dualities to other types of gauge field, and in
particular to the graviton.
The starting point will be to 
generalise the duality transformations of the free vector field to
dualities for free gravitons, governed by 
the linearised Einstein equations, resulting in dual formulations of
linearised gravity in terms of exotic higher-rank gauge fields that were
discussed in physical gauge in [\cst], and motivating the study of 
such higher-rank gauge fields.
In $D=4$, this leads to an $SL(2)$ duality symmetry, just as for the spin
one case.

These properties of the free graviton theory
 cannot be extended to give local covariant dual formulations
of the generally covariant interacting field theory, just as the Maxwell
dualities do not extend to classical Yang-Mills field theory.
In the latter case, the Yang-Mills field theory does not give the full
picture, and at least in certain supersymmetric theories, presumably
should be replaced by some   interacting theory which
does have non-abelian dualities.
It is possible that for gravity too, Einstein or supergravity field
theories do not give the full picture and in fact arise only as a limit
of some interacting theory which enjoys similar duality properties to the
free theory. Some suggestions that this might be the case have arisen in recent 
work on string theory and M-theory [\cst,\csto].
Such dualities would have many implications for our understanding of
gravity, and it seems worthwhile to explore this possibility and look for
evidence in favour of it, or which could rule it out.

Such gravitational dualities should play a role in the interacting theory at
least  for spacetimes with isometries.
Dimensionally reducing $D$-dimensional Einstein gravity on a circle gives an
abelian vector gauge field $A$ in $D-1$ dimensions, the graviphoton.
This can be dualised to a $D-4$ form gauge field $\ti A$ in  $D-1$
dimensions, which couples to magnetically charged states
arising from Kaluza-Klein monople solutions in $D$ dimensions.
It will be shown that this electromagnetic duality of the graviphoton
 can be formulated
in terms of the
$D$-dimensional gravitational field, with some of the dualities
of the free theory extending to dualities of the interacting theory in
the presence of a Killing vector, involving local covariant
transformations of the curvature tensor.
Dimensional reduction on a space with non-abelian isometry group, such as
a sphere, gives rise to non-abelian gauge fields, and if there were 
non-abelian  dualities of the type discussed above that involve these,
these should extend to
dualities of some of the components of the gravitational field
that cannot be local and covariant.

\chapter{Duality in Physical Gauge}


 Dualities are easily derived in physical (light-cone) gauge [\cst].
Introducing transverse coordinates
$i,j=1,...,D-2$, a vector field $A_\mm$
has physical light-cone gauge degrees of freedom $A_i$ in
the vector  representation
of the little group $SO(D-2)$ and can be dualised
to an $n=D-3$ form
$$\ti A_{j_1...j_n}= \ee _{j_1...j_n i} A^i
\eqn\abc$$
The physical degrees of freedom
are in equivalent representations of the little group and so $A$
and
$\ti A$ give
equivalent field theory representations of the  physical degrees
of
freedom.
Covariance suggests that the covariant field giving rise to
the physical degrees of freedom $\ti A_{j_1...j_n}$ should be\
an $n$-form field $\ti A_{\mm_1...\mm_n}$ in $D$ dimensions, and this
should have gauge invariances sufficient to remove the unphysical degrees
of freedom. Then $\ti A$ 
 should be
an $n$-form gauge field, with gauge invariance
$\delta\tilde A=d \lambda$ and field strength $\ti F= d\ti A$.

In the covariant theory, the 
duality relation can be recast as
$$  \ti F \equiv *F
\eqn\dhadfh$$
which is a local covariant relation for the field strengths but not for
the gauge potentials $A_\mm,\ti A_{\mm_1\dots \mm_n}$.
The Maxwell equations in $D$ dimensions for the 2-form field strength 
$F$ with general sources are
$$ dF=*\ti J,\qq d*F= *J
\eqn\maxw$$
for  electric and magnetic currents $J,\ti J$ satisfying the 
conservation laws
$$d*J=0,\qq d*\ti J=0
\eqn\abc$$
with a magnetic source $\ti J$ for the \lq Bianchi 
identity' $dF=0$.
These equations can be recast in a dual form in terms of 
the dual $(D-2)$-form field strength \dhadfh\
as
$$ d\ti F=* J,\qq d*\ti F= *\ti J
\eqn\abc$$
interchanging 
electric and magnetic currents and fields, and interchanging
field equations and Bianchi identities. In regions in which $\ti J=0$, $F$ can be
solved for in terms of a potential $A$ with $F=dA$, while in regions in which
$J=0$ one can solve for $\ti F=d\ti A$ in terms of a potential 
$\ti A$.

In four dimensions, 
$F$ and $\ti F$ are both 2-forms so that  
the transformation $F \to *F$ preserves the form of the field 
equations and is the electromagnetic duality symmetry 
of the Maxwell equations.
In certain theories, it is part of a larger  $SL(2)$ S-duality symmetry under which 
$(F,\ti F)$ transform as a doublet, as do 
the currents $(J,\ti J)$.

A physical gauge graviton in $D$ dimensions is a transverse traceless
tensor
$h_{ij}$ satisfying
$$
h_{ij}=h_{ji}, \qq h_i{}^i=0
\eqn\hid$$
One or both of the indices on $h_{ij}$ can then be replaced by
$n$ anti-symmetric indices to give a dual form.
Dualising one index gives a field 
$$
D_{i_1\dots i_n\, k} = \ee _{i_1\dots i_n l} h^l{}_k
\eqn\abc$$
so that \hid\ implies the
conditions
$$ D_{i_1\dots i_n\, k}=D_{[i_1\dots i_n]\, k}, \qq D_{i_1\dots
i_{n-1}j}{}^j=0, \qq D_{[i_1\dots i_n\, k]}=0
\eqn\abc$$
Dualising on both indices gives a field
$$C_{i_1\dots i_n \, j_1\dots j_n} =\ee_{i_1\dots i_n m} \ee_{ j_1\dots
j_n n}h^{mn}
\eqn\abc$$
which then satisfies
$$C_{i_1\dots i_n j_1\dots j_n}=C_{[i_1\dots i_n ] [j_1\dots
j_n]} =C_{ j_1\dots j_n i_1\dots i_n} 
\eqn\abc$$
together with
$$  
C_{[i_1\dots i_n j_1]j_2\dots j_n}=
0, \qq
\dd^{i_nj_n}C_{i_1\dots i_n j_1\dots j_n}=
 0
\eqn\abc$$
The fields $D_{i_1\dots i_n\, k},C_{i_1\dots i_n j_1\dots j_n}$
should arise from covariant gauge fields
$D_{\mm_1\dots \mm_n\, \nn},C_{\mm_1\dots \mm_n \nn_1\dots \nn_n}$
in $D$ dimensions with gauge symmetries sufficient to remove all but the
desired physical degrees of freedom.
In the next section, these symmetries will be found and the covariant
form of these duality transformations on the field strengths will be
given.

These duality transformations can be generalised to the interacting
theories in physical gauge.
 Yang-Mills theory can be formulated in 
 the light-cone gauge in terms of a Lie algebra valued
transverse vector potential $A_i$, and it is straightforward to make the 
substitution
$$ A_i={1\over n!} \ee _{ij_1...j_n}\ti A^{j_1...j_n}
\ek
to formulate the theory in terms
of a Lie algebra valued transverse $n$-form $\ti A$.
As the substitution is local, the resulting theory is a {\it local}
intreracting theory of $\ti A$, although there are the usual light-cone
gauge features such as inverse powers of the longitudinal momentum $p^+$.
However, there is a problem in finding a covariant gauge theory 
that gives rise to this theory on going to  physical 
gauge. The theory written in terms of $A$ does arise from gauge-fixing
the standard covariant Yang-Mills theory, but there is no local covariant
classical field theory of a  Lie algebra valued 
 $n$-form gauge field with suitable gauge invariances that gives rise to
this physical gauge theory of $\ti A$ on gauge-fixing.

For gravity, the full interacting Einstein theory can be given in
light-cone gauge as a theory of a symmetric transverse traceless
tensor field
$h_{ij}$
with non-polynomial interactions [\gor]. (The field $h_{ij}$ parameterises the coset
space  $SL(D-2)/SO(D-2)$, and this non-linear sigma-model structure is the origin of
many
  of the non-polynomial interactions [\gor].)
 Here one can simply make the local substitution
$$
h_{kl}
 = {1\over n!}\ee ^{i_1\dots i_n  }{}_l D_{i_1\dots i_n\, k}
\eqn\abc$$
or
$$h_{mn} ={1\over m!n!}\ee^{i_1\dots i_n  }{}_m \ee^{
j_1\dots j_n  }{}_n 
C_{i_1\dots i_n \, j_1\dots j_n}
\eqn\abc$$
to obtain an interacting  physical gauge theory of the dual
potentials
$D$ or $C$, whose only non-localities are those involving
inverse powers of the longitudinal momentum $p^+$.
 Again, while the theory
written in terms of
$h_{ij}$ arises from the gauge fixing of Einstein's theory, 
the
construction of a   local covariant classical gauge theory of $C$ or $D$ that gives
rise to these theories on gauge-fixing   appears   to be problematic.

In each case, there is no problem in dualising the interacting physical
gauge theory and writing it as a local theory of dual potentials.
In the free case, one can find covariant interacting theories that give
rise to each of these dual forms, but in the interacting case, it appears that only 
some of these dual forms of the physical gauge theories arise from gauge
fixing local covariant theories.


\chapter{Linearised Gravity}

A free 
symmetric tensor gauge field $h_{\mm\nn}$ in $D$ dimensions has  
the gauge symmetry 
$$ \dd h_{\mm\nn}= \pa _{(\mm}\xi_{\nn)}
\eqn\abc$$
and the invariant field strength is the linearised Riemann tensor
$$ R_{\mm\nn\, \ss\tt}=\pa_{\mm} \pa_{\ss}h_{\nn\tt}+\ldots=-4\pa_{[\mm}h_{\nn][\ss,\tt]}
\eqn\ris$$
This satisfies 
$$R_{\mm\nn\, \ss\tt}=R_{\ss\tt\,\mm\nn}
\eqn\abc$$
together with
the first Bianchi identity
$$R_{[\mm\nn\, \ss]\tt} =0
\eqn\rbo$$
and the second Bianchi identity
$$\pa _{[\rr}R_{\mm\nn]\, \ss\tt}=0
\eqn\rbot$$
The natural free field equation  in $D\ge 4$ is the linearised 
Einstein equation
$$R^\ss{}_{\mm\, \ss\nn}=0
\eqn\rfo$$
 which together with \rbo\ implies 
$$\pa ^\mm R_{\mm\nn\, \ss\tt}=0
\eqn\rft$$
where indices are raised and lowered with a flat background metric 
$\eta_{\mm\nn}$.

Dualising the linearised curvature gives tensors
$S=*R$ and $G=*R*$ with components
$$S_{\mm_{1}\mm_{2}\ldots\mm_{n+1}\;
\nu\rr }
=\2
\ee _{\mm_{1}\mm_{2}\ldots\mm_{n+1}\aa\bb}
R^{\aa\bb }{}_{\nu\rr}
\eqn\sis$$
and
$$G_{\mm_{1}\mm_{2}\ldots\mm_{n+1}\;
\nu_{1}\nu_{2}\ldots\nu_{n+1} }
=\4
\ee _{\mm_{1}\mm_{2}\ldots\mm_{n+1}\aa\bb}
\ee_
{\nu_{1}\nu_{2}\ldots\nu_{n+1} \gg \dd}R^{\aa\bb\gg\dd}
\eqn\gis$$
where
$$n=D-3
\eqn\abc$$

The conditions \rbo-\rft\ then become 
the equations for $G$ given respectively by
$$G_{[\mm_{1}\mm_{2}\ldots\mm_{n+1}\,
\nu_{1}]\nu_{2}\ldots\nu_{n+1} }
=0
\eqn\gbo$$
{}
$$\pa^{\mm_{1}}G_{\mm_{1}\mm_{2}\ldots\mm_{n+1}\,
\nu_{1}\nu_{2}\ldots\nu_{n+1} }
=0
\eqn\gft$$
{}
$$G_{\nn \mm_{1}\mm_{2}\ldots \mm_{n}\, \rr}{} ^{\mm_{1}\mm_{2}\ldots \mm_{n}
  } 
=0
\eqn\gfo$$
{}
$$\pa_{[\rr}G_{\mm_{1}\mm_{2}\ldots\mm_{n+1}]\,
\nu_{1}\nu_{2}\ldots\nu_{n+1} }
=0
\eqn\gbt$$
Here  \gbo\ can be regarded as a first Bianchi identity and 
\gbt\ can be regarded as a second
Bianchi identity, implying that
$G$ can be solved for in terms of a potential
$$
C_{\mm_{1}\mm_{2}\ldots\mm_{n}\,
\nu_{1}\nu_{2}\ldots\nu_{n} }
=
C_{[\mm_{1}\mm_{2}\ldots\mm_{n}]\,
[\nu_{1}\nu_{2}\ldots\nu_{n}] }
\eqn\abc$$
satisfying 
$$
C_{[\mm_{1}\mm_{2}\ldots\mm_{n}\,
\nu_{1}]\nu_{2}\ldots\nu_{n} }
=0
\eqn\abc$$
by the expression
$$G_{\mm_{1}\mm_{2}\ldots\mm_{n+1}\,
\nu_{1}\nu_{2}\ldots\nu_{n+1} }
=\pa_{[\mm_{1}}C_{\mm_{2}\ldots\mm_{n+1}]\,[
\nu_{1}\nu_{2}\ldots \nu_{n},\nu_{n+1}] }
\eqn\abc$$
The field strength is invariant under the
gauge transformations
$$ \dd C_{\mm\nn\ldots\kk , \rr\ss\ldots\ll} = \pa _{[\mm}
\chi _{\nn\ldots\kk ]\rr\ss\ldots\ll} +\pa _{[\rr} \chi
_{\ss\ldots\ll]\mm\nn\ldots\kk}-2
\pa _{[\mm} \chi _{\nn\ldots\kk\rr\ss\ldots\ll]}
\eqn\delcis$$
with parameter 
$$\chi _{\mm_1\ldots \mm_{n-1}\nn_1 \ldots\nn _n}=
\chi _{[\mm_1\ldots \mm_{n-1}][\nn_1 \ldots\nn _n]}
\eqn\abc$$
Then \gfo\ can be regarded as the field equation, implying \gbt.


Similarly, the conditions \rbo-\rft\   become 
the equations for $S$ given respectively by
$$S_{\mm_{1}\mm_{2}\ldots\mm_{n}\rr\,
\nu }{}^{\rr}=
0\eqn\sfo$$
{}
$$\pa^{\ss} S_{\ss\mm_{1}\mm_{2}\ldots\mm_{n}\,
\nu\rr }=
0, \qq
S_{\mm_{1}\mm_{2}\ldots\mm_{n+1}\,
[\nu  \rr,\ss]}=
0
\eqn\sft$$
{}
$$S_{[\mm_{1}\mm_{2}\ldots\mm_{n+1}\,
\nu]\rr }=0,\qq 
S_{\ss[\mm_{1}\mm_{2}\ldots\mm_{n}\,
\nu  \rr]}=
0
\eqn\sbo$$
{}
$$\pa_{[\ss}S_{\mm_{1}\mm_{2}\ldots\mm_{n+1}]\,
\nu\rr }=
0, \qq \pa^{\rr}
S_{\mm_{1}\mm_{2}\ldots\mm_{n+1}\,
\nu\rr }=0
\eqn\sbt$$
Here  \sbo\ can be regarded as first Bianchi identities and the first 
equation in \sbt\ and the second equation in \sfo\ can be regarded as second
Bianchi identities, implying that
$S$ can be solved for in terms of a gauge field
$$D_{\mm_{1}\mm_{2}\ldots\mm_{n}\, \nu}
=D_{[\mm_{1}\mm_{2}\ldots\mm_{n}]\, \nu}
\eqn\abc$$
satisfying
$$D_{[\mm_{1}\mm_{2}\ldots\mm_{n}\, \nu]}
=0\eqn\abc$$
with 
$$S_{\mm\nn\ldots \rr\; \ss\tt} =\pa _{[\mm} D_{\nn\ldots \rr]\,  [\ss,\tt]}
\eqn\abc$$
The field strength 
$S$ is then invariant under the gauge transformation
$$\eqalign{ \dd
D_{\mm \nn \ldots \ss \, \rr}
&= \pa _{[\mm} \aa _{\nn\ldots \ss]\rr} 
\cr &+ \pa _ \rr \bb _{\mm \nn \ldots \ss}-\pa _ {[\rr} \bb _{\mm \nn
\ldots
\ss]}
\cr}
\eqn\abc$$
with parameters 
$$\aa _{\mm_1\ldots \mm_n \rr}= \aa _{[\mm_1\ldots \mm_n]\rr},\qq
\aa _{[\mm_1\ldots \mm_n \rr]}=0,
\qq  \bb _{\mm \nn \ldots \ss}= \bb _{[\mm \nn \ldots \ss]}
\eqn\abc$$
Then \sfo\ can be taken as the field equation for the gauge field
$D$, which then implies the first equation in \sft\ and the second in 
\sbt.
Such gauge fields were first considered in [\curt], developing the discussion of  
massive gauge fields of [\CF],  and further discussed in [\AKO-\KO].

In $D=5$, $ C_{\mm\nn\, \rr\ss}$ has the algebraic properties of the 
Riemann tensor; such gauge fields 
played a special role in [\cst] and similar gauge fields in $D=4$  
were considered in [\notivarg,\csto].
In $D=4$, all three dual gauge fields 
$h_{\mm\nn},D_{\mm\nn},C_{\mm\nn}$ are symmetric tensor gauge  fields
  (linearised gravitons)
with curvatures $R_{\mm\nn \rr\ss},S_{\mm\nn \rr\ss},G_{\mm\nn 
\rr\ss}$.
In $D=3$, the Weyl tensor vanishes identically and the Riemann 
tensor is completely determined by the Ricci tensor, so that the field 
equation \rfo\ implies that 
the field strength \ris\ vanishes and $h_{{\mm\nn}}$ is pure gauge.
The simplest non-trivial linear field equation 
in $D=3$ is
$$R^{\mm\nn}{}_{\mm\nn}=0
\eqn\rfot$$
representing one degree of freedom. 
The curvature can then be dualised to give
$G_{\mm\nn}$ ($G=*R*$) 
satisfying \gbo,\gbt,\gft\ but with \gfo\ replaced by the field equation
$$G_{\mm}{}^{\mm}=0
\eqn\gfot$$
which follows from \rfot.
The potential $C$ is a scalar
and
$$ G_{\mm\nn}=
\pa_{\mm}\pa_{\nn}C
\eqn\abc$$
so that \gfot\ implies the scalar satisfies the free scalar field equation
$$
\pa ^{2}C=0
\eqn\abc$$
Similarly, the curvature can be dualised to
$S_{\mm\, \nn_{1}\nn_{2}}$ ($S=*R$)
satisfying \sfo,\sbt,\sft\ but with \sbo\ replaced by 
$$
S_{[\mm\, \nn_{1}\nn_{2}]}=0
\eqn\abc$$
Then $S$ can be solved for in terms of a vector potential $D_{\mm}$
with
$$S_{\mm\, \nn_{1}\nn_{2}}
=\pa_{\mm}\pa_{[\nn_{1}}D_{\nn_{2}]}
\eqn\abc$$
and \sfo\ is the Maxwell equation
$$\pa^{\mm}\pa_{[\mm}D_{\nn]}=0
\eqn\abc$$
The $D=3$ duality betweeen a vector field $D_{\mm}$ and a scalar $C$ 
is a well-known example of the electromagnetic duality of section 2, 
but here it is seen that they are also dual to a free graviton with 
the field equation \rfot.
In $D=2$, the  Riemann 
tensor is completely determined by the Ricci scalar, so that the field 
equation \rfot\ only has trivial solutions 
and there is no non-trivial linear field equation involving 
$h_{\mm\nn}$ alone.

\chapter{General Tensor Gauge fields}

It has been seen that a symmetric tensor gauge field can be dualised on
one or both indices to get a new field representation of the same degrees
of freedom. In this section, this will be generalised to other tensor
gauge fields, which in principle can be dualised on any subset of their
indices. The strategy is to start with a light-cone gauge potential, dualise
to a  potential in an equivalent representation of the little group, then seek
a covariant formulation based on the dual potential. A general method for
constructing a covariant field theory from any free light-cone gauge theory
with a potential in any represetnation of the little group was given in
[\Sieg].

Consider a gauge field with physical degrees of freedom
$$D_{i_1\dots i_r \, j_1\dots j_s} =
D_{[i_1\dots i_r] \, [j_1\dots j_s]}
\eqn\abc$$
represented by a Young tableau with two columns, one of length $r$ and one of length $s$.
This   satisfies
$$  
D_{[i_1\dots i_r j_1]j_2\dots j_s}=
0, \qq
D_{i_1\dots i_{r-1}[i_r \, j_1\dots j_s]} =0
\qq
\dd^{i_rj_s}
D_{i_1\dots i_r \, j_1\dots j_s} =
 0
\eqn\abc$$
It will be convenient to refer to an element of
$\lll ^r \otimes \lll ^s \otimes \dots \otimes \lll^t $
(where $\lll ^r $ is the space of $r$-forms) 
represented by a Young tableau with columns of length $r,s,...,t$
as a form of type
$[r,s,\dots, t]$, so that
$D_{i_1\dots i_r \, j_1\dots j_s} $ is an $[r,s]$ form.
This can be dualised on the first $r$ indices to give a dual form of type
$[\ti r,s]$, on the second set of indices to 
give an $[r,\ti s]$ form, or on both sets of indices to give 
an $[\ti r, \ti s]$ form, where $\ti r= n+1-r$, 
$\ti s= n+1-s$. (If $\ti r<s$, it is conventional to 
take the longest column first, so that    strictly speaking this is a 
tableau with the first column of  length $s$ and the second of length $\ti r$. Such
re-orderings are to be understood where necessary in what follows.)

For example, dualising on the first set
gives
$$\ti D_{k_1\dots k_{\ti r} \, j_1\dots j_s}={1\over r!}\ee _{k_1\dots k_{\ti r}}{}^{
i_1\dots i_r} D_{i_1\dots i_r \, j_1\dots j_s}\ek

The physical degrees of freedom given by an $[r,s]$ form of $SO(D-2)$
come from a covariant gauge field 
$$D_{\mm_1\dots \mm_r \, \nn_1\dots  \nn_s} =
D_{[\mm_1\dots \mm_r] \, [ \nn_1\dots  \nn_s]}
\eqn\abc$$
which is an $[r,s]$ form of the Lorentz group $SO(D-1,1)$
 satisfying
$$D_{[\mm_1\dots \mm_r \, \nn_1]\nn_2\dots  \nn_s} =0,\qq
D_{\mm_1\dots \mm_{r-1}[\mm_r \, \nn_1\dots  \nn_s]} =
0
\eqn\abc$$
The gauge transformations for such a gauge field are [\Morr]
$$
\dd D_{\mm_1\dots \mm_r \, \nn_1\dots  \nn_s}
= P_{r,s}\left[
\pa_{[\mm_1}\aa_{\mm_2\dots \mm_r] \, \nn_1\dots  \nn_s}
+\bb _{\mm_1\dots \mm_r \, [\nn_1\dots \nn_{s-1}, \nn_s ]}
\right]
\eqn\abc$$
where $\aa$ is a form of type $[r-1,s]$,  $\bb$ is a form of type $[r ,s-1]$
and $P_{r,s}$ is the projector onto forms of type $[r,s]$ (Young symmetriser).
These gauge transformations preserve the field strength
$$S_{\mm_1\dots \mm_{r+1} \, \nn_1\dots  \nn_{s+1}}
=- \pa _{[\mm_1}D_{\mm_2\dots \mm_{r+1}] \, [\nn_1\dots  \nn_s,\nn_{s+1}]}
\eqn\abc$$
The natural linear  free field equations in $D\ge r+s+2$ are
$$S_{\mm_1\dots \mm_{r+1} \, \nn_1\dots  \nn_{s+1}}\eta ^{\mm_1\nn_1}
=0
\eqn\abc$$
However, in dimension $D=r+s+1$, these field equations imply that the 
field strength vanishes identically, as was seen in particular examples
in the previous section, and the simplest non-trivial  linear field
equation is
$$S_{\mm_1\dots \mm_{r+1} \, \nn_1\dots  \nn_{s+1}}\eta ^{\mm_1\nn_1}\eta
^{\mm_2\nn_2} =0
\eqn\abc$$
with two contractions.
In $D=r+s$, this equation only has trivial solutions, and the 
 simplest non-trivial 
linear field equation is
$$S_{\mm_1\dots \mm_{r+1} \, \nn_1\dots  \nn_{s+1}}\eta ^{\mm_1\nn_1}\eta
^{\mm_2\nn_2}\eta ^{\mm_3\nn_3} =0
\eqn\abc$$
with three contractions. Similarly, the
field equation in $D=r+s+2-p$ for $p\le r,p\le s$ is 
$$S_{\mm_1\dots \mm_{r+1} \, \nn_1\dots  \nn_{s+1}}\eta ^{\mm_1\nn_1}
\dots \eta ^{\mm_p\nn_p} =0
\eqn\abc$$
with $p$ contractions.

In the covariant gauge theory, the duality transformations are 
given in terms of the field strengths.
An $[r,s]$ gauge field $D_{\mm_1\dots \mm_r \, \nn_1\dots  \nn_s} $
 is dualised to a $[\ti r,s]$ gauge field $\ti D_{\mm_1\dots \mm_{\ti r} \, \nn_1\dots 
\nn_s}
$. The relation between the gauge fields is non-local, but the
relation between the $[r+1,s+1]$ field strength $S$ for $D$  and the 
 $[\ti r+1,s+1]$ field strength $\ti S$ for $\ti D$ 
is local,
$\ti S=*S$.
The field strengths for the $[\ti r,s]$, $[ r,\ti s]$   and $[\ti r,\ti s]$  gauge fields
dual to $D$ are respectively $*S,S*$ and $*S*$.

This extends straightforwardly to gauge fields which are
$[r,s,\dots , t]$ forms, which can be dualised
on any set of anti-symmetric indices.
For example, an $[r,s,t]$ form can be dualised 
to forms of types $[\ti r,s,t]$, $[r,\ti s,t]$, $[r,s,\ti t]$, 
$[\ti r,\ti s,t]$, $[r,\ti s,\ti t]$, $[\ti r,s,\ti t]$, $[\ti r,\ti
s,\ti t]$.
The field strength of a gauge field of type $[r_1,r_2,\dots, r_m]$
is given by acting on the gauge field with $m$ derivatives and is a form
of type
$[r_1+1,r_2+1,\dots, r_m+1]$.

Consider next a 2-form gauge field $B_{\mm\nn}$ with gauge invariance
$\dd B=d \ll$. This has physical degrees of freedom given by
a transverse 2-form $B_{ij}$, which can be dualised in the usual way to give an
$n-1$ form 
$$\ti B_{i_1\dots i_{n-1}}=\2
\ee _{i_1\dots i_{n-1}pq }
B^{pq}\eqn\btiis$$
arising from gauge-fixing an 
$n-1$ form  gauge field.
Instead, one could attempt to dualise the anti-symmetric tensor in the same
way as was done in the last section for a symmetric tensor.
Dualising   on one index
gives a tensor
$$\hat B_{i_1\dots i_n j }
=\ee _{i_1\dots i_{n}  }{}^k
B_{kj}
\eqn\abc$$
However, the antisymmetry of $B_{ij}$ implies that $\hat B$ is pure trace
(i.e. the trace-free part vanishes) so that
$$\hat B_{i_1\dots i_n j }
=\ti B_{[i_1\dots i_{n-1}}\dd_{i_n]j}
\eqn\abc$$
and the usual dual $\ti B$ defined by \btiis\ is recovered.
Similarly, dualising on both indices
gives
$$\bar B_{i_1\dots i_n j_1\dots j_n}=
\ee _{i_1\dots i_{n}  }{}^k
\ee _{j_1\dots j_{n}  }{}^l
B_{kl}
\eqn\abc$$
but this is necessarily of the form
$$\bar B_{i_1\dots i_n j_1\dots j_n}=
\2 (\ee _{i_1\dots i_{n} [j_1 }\ti B_{j_2\dots j_{n}]}
+\ee _{j_1\dots j_{n} [i_1 }\ti B_{i_2\dots i_{n}]})
\eqn\abc$$
and again the usual dual $\ti B$ is recovered.
This generalises  to an $r$ form gauge field, which can be dualised 
in the usual way to an antisymmetric tensor gauge field of rank $\ti
r=n+1-r$, and again attempting to dualise on a subset of the indices
leads back to the standard $\ti r$-form dual.

A general second rank tensor
$k_{ij}$  can be decomposed into symmetric, anti-symmetric and
trace parts
$$k_{ij}=h_{ij}+B_{ij} + \dd_{ij} \phi
\eqn\abc$$
arising from a graviton $h_{\mm\nn}$, a 2-form gauge field $B_{\mm\nn}$
and a scalar $\phi$.
This can be dualised on the first index
to give a    tensor
$$\hat k_{i_1\dots i_n j }
=\ee _{i_1\dots i_{n}  }{}^k
k_{kj}
\eqn\abc$$
which is anti-symmetric on the first $n$ indices 
$\hat k_{i_1\dots i_n j }=\hat k_{[i_1\dots i_n] j }$
but is otherwise
arbitrary.
This decomposes into
 $$\hat k_{i_1\dots i_n j }=D_{i_1\dots i_n j }+\ti B_{[i_1\dots
i_{n-1}}\dd_{i_n]j}+\ee _{i_1\dots i_{n}  j}
\phi
\eqn\abc$$
where $D_{i_1\dots i_n j }$, $ \ti B_{i_1\dots
i_{n-1}}$ are the duals of 
$h_{ij},B_{ij}$  obtained above.
Similarly, dualising on both indices
gives
$$\bar k_{i_1\dots i_n j_1\dots j_n}=
\ee _{i_1\dots i_{n}  }{}^k
\ee _{j_1\dots j_{n}  }{}^l
k_{kl}
\eqn\abc$$
which is given in terms of the duals $C_{i_1\dots i_n j_1\dots j_n}$, 
 $ \ti B_{i_1\dots
i_{n-1}}$ 
as
$$\eqalign{\bar k_{i_1\dots i_n j_1\dots j_n}=&
C_{i_1\dots i_n j_1\dots j_n}
\cr &
+\2 
(\ee _{i_1\dots i_{n} [j_1 }\ti B_{j_2\dots j_{n}]}
+\ee _{j_1\dots j_{n} [i_1 }\ti B_{i_2\dots i_{n}]})
\cr &
+ \phi \ee _{i_1\dots i_{n} k }\ee _{j_1\dots j_{n}}{}^k
\cr}
\eqn\abc$$


The general situation can be summarised as follows.
Consider some tensor gauge field $B_{\mm\nn\dots \rr}$ in $D$ dimensions
whose physical degrees of freedom are represented in light cone gauge by
a  field $B_{ij\dots k}$ in some tensor representation of the little group
$SO(D-2)$. 
Without loss of generality, attention can be restricted to irreducible
representations, as the general case can be decomposed in terms of these,
and each can be considered separately. Then any one of the indices
$ij\dots k$ can be dualised to give
$n=D-3$ antisymmetric indices and dual fields
$\hat B_{i_1\dots i_n j\dots k}=\hat B_{[i_1\dots i_n] j\dots k}$,
$\bar B_{i j_1\dots j_n\dots k}=\bar B_{i [j_1\dots j_n]\dots k}$ and so
on. Any number of indices can be dualised simultaneously; dualising the
first two indices, for example, will give a gauge field
 $b_{i_1\dots i_n  j_1\dots j_n\dots k}$.
Similarly, a set of $r$ antisymmetrised indices can be dualised to
a set of  $\ti r = n+1-r$ antisymmetrised indices.
Not all the dual forms obtained in this way are independent.
In general, the gauge field will be in an irreducible representation 
corresponding to a Young tableau with
$M$ columns of heights $r_a$, $a=1,\dots ,M$, and the independent dual
representations are given by  chosing any set of   columns and 
replacing them by columns of dual height $\ti r_a =n+1-r_a$.
Finally, for each of the dual forms of the tensor    field in physical
gauge, one can construct a covariant gauge field and its  gauge invariances,
using the methods of [\Sieg],   that would lead  to that physical gauge field,
and give a covariant form of the duality relations. 
The covariant gauge potential would typically be 
in the representation of $SO(D-1,1)$ represented by a Young tableau with
columns of the same lengths $r_a$ as for the light-cone gauge potential, with
the possibility of adding an arbitrary number of further columns, all of
length $D-2$ [\pri].
   The relations between the covariant gauge
potentials is non-local, but there is a local relation between dual field
strengths. Further details will be given elsewhere.

\chapter {Sources}

In Maxwell theory, one can introduce an electric source $J_{\mm}$
which couples to $A_{\mm}$ and an $n=D-3$ form magnetic current 
$J_{\mm_{1}\mm_{2}\ldots \mm_{n}}$ which couples to
the dual potential $\ti A_{\mm_{1}\mm_{2}\ldots \mm_{n}}$, with the 
field equations \maxw.
In regions in which the magnetic current $\ti J$ vanishes, $F$ can be solved for in terms 
of a potential $A$, $F=dA$, while
in regions in which the
electric current $ J$ vanishes, $\ti F$ can be solved for in terms 
of a dual potential $\ti A$, $\ti F=d\ti A$.

In linearised gravity, one can   consider adding sources
$T_{\mm\nn}$, $U _{\mm_{1}\mm_{2}\ldots \mm_{n}\, \nn}$
and $V_{\mm_{1}\mm_{2}\ldots \mm_{n} \nn_{1}\nn_{2}\ldots \nn_{n}}$
coupling naturally to 
the potentials
$h_{\mm\nn}$, $D _{\mm_{1}\mm_{2}\ldots \mm_{n}\, \nn}$
and $C_{\mm_{1}\mm_{2}\ldots \mm_{n} \nn_{1}\nn_{2}\ldots \nn_{n}}$, 
respectively.
The linearised Einstein equations (for $D>2$) are
$$R^\ss{}_{\mm\, \ss\nn}=\bar T_{\mm\nn}
\eqn\rfoss$$
where 
$$ \bar T_{\mm\nn}=  T_{\mm\nn}+{1\over D-2} \eta_{\mm\nn}T
\eqn\abc$$
and $ T_{\mm\nn}$ is the energy-momentum tensor, the trace term 
being added so that the Bianchi identities imply that
$ T_{\mm\nn}$ is identically conserved,
$\pa ^{\mm} T_{\mm\nn}=0$.
Similarly, the natural field equations with the source 
$U $ is
$$S_{\mm_{1}\mm_{2}\ldots\mm_{n}\rr\,
\nu }{}^{\rr}=\bar U_{\mm_{1}\mm_{2}\ldots\mm_{n} \,
\nu } 
\eqn\sfoss$$
where
$$\bar U_{\mm_{1}\mm_{2}\ldots\mm_{n} \,
\nu } =  U_{\mm_{1}\mm_{2}\ldots\mm_{n} \,
\nu } +{n\over 2}
 \eta _{\nn [\mm_{1}} U_{\mm_{2}\mm_{3}\ldots\mm_{n}] \rho }{}^{\rho}
 \eqn\abc$$
 so that the Bianchi identities for $S$ imply that $U$ is conserved,
 $$\pa ^{\mm_{1}} U_{\mm_{1}\mm_{2}\ldots\mm_{n} \,
\nu } =0, \qq\pa^{\nn}
 U_{\mm_{1}\mm_{2}\ldots\mm_{n} \,
\nu } =0
\eqn\abc$$
The field equation \sfoss\ then implies that $U$ is a source for the 
gravitational Bianchi identity \rbo:
$$R_{[\mm\nn\, \ss]\tt}
={1\over n!}
\ee _{\mm\nn\, \ss}{}^{\mm_{1}\mm_{2}\ldots\mm_{n}}
\bar U_{\mm_{1}\mm_{2}\ldots\mm_{n} \,
\tt } 
\eqn\rbossa$$
Similarly, the stress-energy tensor $T_{\mm\nn}$ is a \lq source' on 
the right hand side of the Bianchi identity \sbo.
In a region in which $U$ vanishes, the Bianchi identity holds and 
the gravitational field can be written in terms of $h_{\mm\nn}$ 
satisfying the field equation \rfoss, while in a region in which $T$ 
vanishes, the Bianchi identities \sbo\ hold and 
the linearised  gravitational field
can be expressed in terms of the dual field $D_{\mm\ldots\nn}$ with 
the field equation \sfoss.

Including a source for the field equation \gfo\
gives 
$$G_{\nn \mm_{1}\mm_{2}\ldots \mm_{n}\, \rr}{} ^{\mm_{1}\mm_{2}\ldots \mm_{n}
  } 
=\bar V_{\nn  \rr}
\eqn\gfoss$$
However, it follows from \gis,\rfoss\ that
$\bar V_{\nn  \rr}$ is related to the usual energy-momentum tensor,
$$\bar V_{\nn  \rr}=aT_{\nn  \rr}+b \eta _{\nn  \rr} T
\eqn\abc$$
for some $n$-dependent coefficients $a,b$.
 A source for \gbo\ of the form
$$G_{[\mm_{1}\mm_{2}\ldots\mm_{n+1}\,
\mm_{n+2}]\nu_{1}\ldots\nu_{n} }
=W_{[\mm_{1}\mm_{2}\ldots\mm_{n+1}\,
\mm_{n+2}]\nu_{1}\ldots\nu_{n} }
\eqn\gbos$$
can be given in terms of $U$, so that
$$G_{[\mm_{1}\mm_{2}\ldots\mm_{n+1}\,
\mm_{n+2}]\nu_{1}\ldots\nu_{n} }=
\ee_{\mm_{1}\mm_{2}\ldots\mm_{n+1}\,
\mm_{n+2}}{}^{\rr}
\bar U _{\nu_{1}\ldots\nu_{n} \, \rr}
\eqn\abc$$
Thus, although linearised gravity can be represented in terms of three
different types of fields, there are only two independent types of source
that can arise, $T$ and $U$.

Consider then linearised gravity, formulated in terms of a field strength
$R_{\mm\nn\rr\ss}$, satisfying these generalised field equations and \lq
Bianchi identities' with sources $T,U$.
If there is no magnetic source, $U=0$, the standard
Bianchi identities hold and the field strength can be solved for in terms
of the gauge field $h_{\mm\nn}$ as usual, or in terms of the double-dual
potential $C$.
 However, in a region in which $U\ne 0$ but $T=0$, one can
instead solve for $R$ in terms of the dual potential $D$
and the theory has to be formulated in terms of this dual potential.
Outside regions in which $U\ne 0$, one can find a linearised metric 
$h_{\mm\nn}$ locally, but the presence of regions with non-zero $U$ will
often lead to Dirac string singularities in $h_{\mm\nn}$.
The situation is similar for the more general fields considered in section 4.

 Some of the possible physical consequences of a magnetic source for
gravity were discussed in [\zee].
It seems unlikely that there is   
such \lq magnetic energy' in the observable universe.
However, while it is expected that there are very few magnetic
monopoles in the observable universe,   or perhaps none at all, magnetic
monopoles have come to play a central role in our theoretical understanding of the
non-perturbative structure of many gauge theories, and it is conceivable
that magnetic sources for gravity could play a similarly important role
in non-perturbative gravity, despite their apparent absence. As will be seen, there is
a sense in which Kaluza-Klein monopoles can be regarded as such magnetic sources.

\chapter{Spacetimes with Killing Vectors}

In this section, the possibility of performing duality 
transformations in the 
non-linear Yang-Mills or Einstein equations will be considered
in the special cases of configurations with an extra symmetry.

In Yang-Mills theory with Lie-algebra valued $A$ and $F=dA+A\wedge A$,
the field equations $D*F=0$ and Bianchi identities $DF=0$ involve the 
background covariant derivative
$D\phi=d\phi +[A, \phi ]$ and the presence of the explicit gauge potential $A$ means 
that there is no  local covariant generalisation of electromagnetic   
duality in general.
However, consider a  gauge-field configuration
that admits a \lq 
Killing scalar' $\aa$ i.e. a gauge field  configuration $A$  that 
is invariant under 
the gauge transformation $\dd A=D\aa$ for some parameter $\aa$, so that
$\aa$ must be covariantly constant
$$D\aa=0\eqn\abc$$
implying $[F,\aa]=0$.
Then the 2-form
$f=tr(\aa F)$
satisfies the Maxwell equations
$$df=0,\qq d*f=0
\eqn\abc$$
and
these can be re-expressed in terms of a dual field strength $\ti 
f=*f$. Furthermore, $f$ can be solved for in terms 
of a potential $a$, $f=da$, and similarly $\ti f$ can be expressed in 
terms of a dual potential, $\ti f =d\ti a$.
Then the existence  of a Killing scalar allows linearisation of the field
equations  for the
components $f$ of $F$ and this subset of the field  equations can be dualised.

Consider a $D$-dimensional spacetime with metric
$g_{\mm\nn}$
and a Killing vector $k^{\mm}$ satisfying the  Killing equation
$$\na _{(\mm}k_{\nn)}=0
\eqn\kill$$
Then the  Killing equation and Ricci identities imply that 
the (full nonlinear) Riemann curvature tensor
satisfies
$$R_{\mm\nn\rr\ss}k^{\ss}=\na _{\rr}f_{\mm\nn}
\eqn\abc$$
where
$$f_{\mm\nn}=2\pa_{[\mm}k_{\nn]}
\eqn\abc$$
The first Bianchi identity 
implies 
$$df=0\eqn\abc$$
while the (full nonlinear) Einstein equation
$$R_{\mm\nn}=\bar T_{\mm\nn},\qq \bar T_{\mm\nn}=  T_{\mm\nn}-{1\over 
D-2} T g_{\mm\nn}
\eqn\abc$$
(where $T=g^{\mm\nn} T_{\mm\nn}$)
implies the Maxwell equation
$$ d*f=*J
\eqn\abc$$
where the current
is
$$J_{\mm}= \bar T_{\mm\nn}k^{\nn}\eqn\abc$$
If $T\ne 0$, this differs from the
conserved momentum
$$P_{\mm}=   T_{\mm\nn}k^{\nn}\eqn\abc$$
(the conservation  of which is a consequence of the isometry symmetry) 
by 
$$J_{\mm}-  P_{\mm}= \2 k_{\mm}R
\eqn\abc$$
which is automatically conserved as  
\kill\ implies $\nabla ^{\mm}(k_{\mm} R)=0$.
Then the presence of a Killing vector has allowed some of Einstein's 
equations -- those  with at least one component in the Killing 
direction -- to be rewritten as the free Maxwell
equations for the field stength $f=da$ with \lq gauge potential' 
$a_{\mm}=g_{\mm\nn}k^{\nn}$.

A magnetic source $\ti J$ for $f$
$$df=*\ti J\eqn\abc$$
would
correspond to a source for the 1st Bianchi identity
$$R_{[\mm\nn\rr]\ss}k^{\ss}= (*\ti J)_{\mm\nn\rr}
\eqn\abc$$
and the presence of such sources would mean that
$f$ couldn't be solved for in terms of a local potential $a_{\mm}$ and
so the curvature couldn't be expressed in terms of a 
metric $g_{\mm\nn}$ alone -- with magnetic monopole sources for $a$, the potential 
$a$ would have Dirac string singularities and so the metric would 
also have these;
such Dirac string singularities in the metric are sometimes referred 
to   as Misner strings.

In the non-linear theory, one can define a tensor $S=*R$, as in \sis.
The Bianchi identities \sft,\sbt\    become
$$\nabla^{\ss} S_{\ss\mm_{1}\mm_{2}\ldots\mm_{n}\,
\nu\rr }=
0, \qq
S_{\mm_{1}\mm_{2}\ldots\mm_{n+1}\,
[\nu  \rr;\ss]}=
0
\eqn\sftaa$$
{}
$$\nabla_{[\ss}S_{\mm_{1}\mm_{2}\ldots\mm_{n+1}]\,
\nu\rr }=
0, \qq \nabla^{\rr}
S_{\mm_{1}\mm_{2}\ldots\mm_{n+1}\,
\nu\rr }=0
\eqn\sbta$$
with covariant derivatives  involving the Christoffel connection. Thus the metric
appears explicitly, and the theory cannot be written in terms of the dual potential
$D$ alone. If there is a magnetic source, the field equation will be \sfoss.
With a Killing vector $k$, $\ti f=*f$
satisfies
$$
\nabla _\rho \ti f _{\mm_{1}\mm_{2}\ldots\mm_{n+1}}=
S_{\mm_{1}\mm_{2}\ldots\mm_{n+1}\, \rho\ss}k^\ss
\ek
In the absence of matter, $T_{\mm\nn}=0$, and $S$ will satisfy \sbo,
so that 
$$ d\ti f=0
\ek
and locally there is a 
dual potential $n$-form
$\ti a$ such that
$$\ti f = d \ti a
\ek
This satisfies the field equation
$$
\nabla ^\rho \ti f _{\mm_{1}\mm_{2}\ldots\mm_{n}\rho}=
 \ti J_{\mm_{1}\mm_{2}\ldots\mm_{n}}
\ek
where
$$ \ti J_{\mm_{1}\mm_{2}\ldots\mm_{n}} = \bar U _{\mm_{1}\mm_{2}\ldots\mm_{n }\,
 \ss}k^\ss
\ek
Thus the presence of a Killing vector allows the rewriting of some of Einstein's
equations in terms of the dual potential $\ti a$. In the linearised theory, this dual
potential is given in terms of the dual gravitational field $D$ by
$$ \ti a _{\mm_{1}\mm_{2}\ldots\mm_{n}}= D  _{\mm_{1}\mm_{2}\ldots\mm_{n }\,
 \ss}k^\ss
\ek

On dimensional reduction from $D$ to $D-1$ dimensions on a circle,
the metric gives rise to an abelian gauge field $A_m$ in $D-1$
dimensions, together with a scalar field $V$ and a metric $g_{mn}$
throught the ansatz
$$ds^2=V(d y + A_m d x^ m)^2 +V^{-1}g_{mn}d x^m
d x^n
\eqn\remet$$
where 
$V,A_m$ and $g_{mn}$ 
depend on the coordinates $x^{m}$, so that $a_\mm=(V,VA_m)$.
The above equations give rise to   standard Kaluza-Klein field equations
for $V,A_m$ and $g_{mn}$.
The vector field   $A_m$
  is defined up to a gauge transformation
$$A_\mm\rightarrow A_m+  \partial_ m \rho
\eqn\varo$$
as such a change  can be absorbed into the coordinate transformation
$$ y \rightarrow y - \rho
\eqn\abc$$
The invariant field-strength
$$F_{mn}= \partial_{[m} A_{n]}
\eqn\twi$$
is the {  twist} or {  vorticity} of the vector field $k$.
If the  potential $A$   has
a Dirac string singularity, then the metric has a Misner 
string singularity.

\chapter{Kaluza-Klein Monopoles and Dimensional Reduction}

A vector field $A_m$ in $d$ dimensions
is associated with electrically charged 0-branes, which couple to $A_m$, and with
magnetically charged $d-4$ branes which couple to the dual
 potential, which is a $d-3$ form $\ti A_{m_1...
m_{d-3}}$.
Suppose that the vector field arises from
 the Kaluza-Klein reduction of gravity from $D$ to $d=D-1$ dimensions on a
circle, with $D\ge 5$. The  $D$-dimensional metric
$g_{\mm\nn}$ of the form \remet\ gives a metric $g_{mn}$, a vector field $A_m$ and a
scalar
$\phi$ in $D-1$
dimensions. 
Then states which are electrically
charged in  $d$-dimensions arise from states 
carrying momentum in the (internal)  $y$ direction, 
with the $D$-momentum $P^\mm$ giving the $D-1$-momentum $P^m$ and the electric
charge
 $Q=P^D$ associated with $A$ on dimensional reduction, $P^\mm=(P^m,Q)$. In particular,
the 0-branes
arise from such momentum modes.
For configurations in $d$ dimensions with magnetic charge,   $A$ will in general have
Dirac string singularities.
These can be avoided by modifying the topology and regarding the potential as a
connection for a non-trivial bundle, or (in the absence of electric charge) by going
to the dual formulation, with  the magnetic charges acting as sources for a
non-trivial dual potential $\ti A$.  
The Dirac string singularities in $A$ give 
rise to Misner string
singularities in
 the $D$ dimensional  metric $g_{\mm\nn}$.
However, in some cases these can be
eliminated by modifying the topology of the $D$ dimensional space time so that it is
not a product of a circle with some space, but is instead a circle bundle.  In
particular, the magnetically charged
$d-4$ branes arise from Kaluza-Klein monopole solutions in
$D$ dimensions of the form $N_4\times \R^{D-5,1}$, where $N_4$ is
Euclidean Taub-NUT space, and the reduction is over the $S^1$ fibre of $N_4$.
For example, the D6-brane of type IIA string theory couples to a RR
7-form potential, the dual of the RR vector field, and arises from the KK monopole
solution of M-theory in $D=11$.
Alternatively, one could attempt to eliminate the string singularities by formulating
the theory in terms of a dual potential, as will now be discussed.

Consider first linearised gravity, with the $D$ dimensional graviton $h_{\mm\nn}$
giving a graviton $h_{mn}$, a vector field $A_m=h_{my}$ and a scalar $\phi = h_{yy}$
in $d=D-1$ dimensions.
In $D$ dimensions, the graviton can be dualised to a gauge field
$D_{\mm_1...\mm_{D-3}\, \nn}$, while in $d$ dimensions, $h_{mn}$
can be dualised to a gauge field
$d_{m_1...m_{d-3}\, n}$, $A_m$
can be dualised to a $d-3$ form gauge field
$\ti A_{m_1...m_{d-3} }$ and $\phi$
can be dualised to a $d-2$ form gauge field
$\ti \phi_{m_1...m_{d-2} }$.
On dimensional reduction, $D_{\mm_1...\mm_{D-3}\, \nn}$
gives rise to the dual graviton $d_{m_1...m_{d-3}\, n}=
D_{m_1...m_{d-3}y\, n}$, the dual scalar
$\ti \phi_{m_1...m_{d-2} }=D_{m_1...m_{D-3} \, y}$
and the dual vector 
$\ti A_{m_1...m_{d-3} }=D_{m_1...m_{d-3}y\, y}$
(the remaining components $D_{m_1...m_{D-3} \, n}$
give no further independent degrees of freedom).
Thus  in the linearised theory, 
magnetically charged branes in $d$ dimensions acting as sources of the gauge field
$\ti A$ lift to configurations that are sources of the components $D_{m_1...m_{d-3}y\,
y}$ of the dual graviton and formulating in terms of $\ti A$ or $D$ avoids the string
singularities that would otherwise arise for $h$ or $A$.
Thus Kaluza-Klein monopoles naturally couple  to the dual graviton $D$, and so can be
thought of as carrying the \lq magnetic energy-momentum' current $U$.

If a BPS  magnetic $d-4$ brane in $d$ dimensions is wrapped on a rectangular $d-4$
torus, then the resulting 0-brane in 4 dimensions has mass proportional to
$$
\ti A_{0m_1...m_{d-4} }R^{m_1}R^{m_2}...R^{m_{d-4}}
\ek
where $R^m$ is the radius of the circle in the $x^m$ direction.
This lifts to the following expression in $D$ dimensions
$$D_{0\mm_1...\mm_{D-4}y\, y}R^{\mm_1}R^{\mm_2}...R^{\mm_{d-4}}R^y R^y
\ek
with the correct quadratic dependence on the radius $R^y$   of the
extra circle.

In the non-linear theory, we have seen  that 
in the presence of a Killing vector, it is possible to dualise some of the
gravitational degrees of freedom, and in particular, it is possible to construct in
this way the dual gravitational potential to which the Kaluza-Klein monopoles  
couple.

\chapter{Duality Symmetries in Four Dimensions}

\section{Maxwell Theory in Four Dimensions}

In four Euclidean dimensions, one can consistently impose the
self-duality condition
$F=*F$
so that the Bianchi identity $dF=0$  implies the field equation $d*F=0$,
but is stronger.
In 3+1 dimensions this is inconsistent as $(*)^{2}=-1$. However, 
given two 2-form field strengths $F_{i}=(F,\ti F)$ 
($i=1,2$)
one can impose
$$F_i= J_i{}^j *F_j
\eqn\fsd$$
for any  matrix $J_i{}^j$ that
  satisfies $J^2=-1$.
  Then   the Bianchi identities $dF_{i}=0$  imply the field equations 
  $d(J_i{}^j*F_{j})=0$, and this is the form of the field equations proposed in 
  [\CJP].
  With the natural choice $$J =\pmatrix{
0 & 1 \cr
-1 & 0\cr
}\eqn\jiss$$
 this gives
  $\ti F=*F$, $F=-*\ti F$.
  More generally, different choices of field-dependent $J$ lead to different field 
  equations and hence define different theories.
  
  The  equations \fsd\
  are  invariant under  an 
    $SL(2,\R)$ symmetry, with $F_{i}$ 
transforming as
$$F \to SF
\eqn\abc$$
where $S_i{}^j$ is an $SL(2,\R)$ matrix
provided  $J$ transforms as 
$$J \to SJS^{-1}
\eqn\jtran$$
For the  standard  theory with coupling constant $g$ and
and a theta-angle $\th$,
a  $J$ with these transformation properties  can be constructed 
as follows.
Combining the coupling constants into the complex modulus
$$ \tau =\tt_1+i\tt_2= \th + {i\over g^2}
\eqn\abc$$ then
  the
flat 2-metric
 $\gamma_{ij}$ 
 given by
 $$ \gamma_{ij}={V\over \tt_2}\pmatrix{
1 & \tau_1 \cr
\tt_1 & |\tt|^2 \cr
}
\eqn\abc$$
transforms as
 $$ \gg \to S\gg S^{t}
\eqn\abc$$
provided $\tt$ transforms under $SL(2)$ through 
fractional linear transformations and the volume $V$ is invariant.
Then introducing 
the  alternating tensor $\epsilon_{ij}$ 
with components $\epsilon_{12}= \sqrt{\gg}$, $\gg=det \gg_{ij}$, a
$J$ transforming as \jtran\ and satisfying $J^{2}=-1$ can be defined as
 $$J_i{}^j
={1 \over \sqrt
{\gamma}}\gamma_{ik}\epsilon^{kj}
\eqn\jis$$
 This is the well-known 
$SL(2,\R)$  duality symmetry of the classical theory which, in the quantum 
theory with $N=4$ supersymmetry,   is broken to a discrete $SL(2,\Z)$
subgroup.   The 2-metric can be interpreted as the metric on a 2-torus, 
with the 
$SL(2,\Z)$ acting as large diffeomorphisms on the 2-torus.

This can be generalised to certain  theories of abelian gauge fields 
interacting with scalars, such as those that occur in the bosonic 
sector of ungauged supergravity 
theories, with $m$ gauge fields $F_{a}$
($a=1,\ldots,m$) whose   field equations can be written
as $d\ti F_{a}=0$, where $\ti F_{a}$ is given in terms of $F_{a}=dA_{a}$
in terms of the variation of the bosonic action $S$ by
$$\ti F _{a}\equiv *(\dd S/\dd F_{a} )= *F_{a} +\ldots\eqn\sdfbf$$
The $2m$ field strengths $F_{i}=(F_{a},\ti F_{a})$ are combined into a
$2m$-vector $F_{i}$ and a scalar-dependent $J$ is determined by 
requiring   that $J^{2}=-1$ and that
\fsd\ implies \sdfbf.
The equations will then be invariant under some duality group $G$ 
for which $F_{i}$ transforms as a $2m$-dimensional representation
and the transformations of the scalars are such that 
$J$ transforms in the appropriate way.
For example, in   $N=4$ supergravity theories, 
each vector field has field strengths fitting into an $SL(2)$ doublet, 
with a $J$ constructed as in \jis\ from
a complex scalar field  $\tt$ taking values in the coset space 
$SL(2,\R)/U(1)$ and transforming as in \jtran.

\section{Linearised Gravity   in Four Dimensions}

Gravitational analogues of electromagnetic duality have been considered in
four-dimensional higher derivative theories of gravity [\Nepo], and in other
gravitational theories in [\Mex], but here attention will be restricted to the
linearised Einstein theory.  In four Euclidean dimensions, one
can impose the self-duality condition $$ R_{\mm\nn\, \ss\tt}=\2
\ee_{\mm\nn\rr\kk}R^{\rr\kk}{}_{\rr\ss} \eqn\abc$$
 or $R=*R$  so that the Bianchi identity \rbo\ implies the field equations 
 \rfo, but with Lorentzian signature this would imply zero curvature.
In 3+1 dimensions, the dual potentials
$D_{\mm\nn}$ and $C_{\mm\nn}$ are again symmetric tensor gauge fields 
and
one can  seek a generalisation of the
construction used above for the Maxwell field.
The curvature tensor 
 $R_{\mm\nn\rr\ss}$ 
 can be dualised on the first two indices
to give $*R$, on the second two to give $R*$ or both to give 
$*R*$.
These can be formed into a $2\times 2 $
matrix of  tensors
$\bar
R_{\mm\nn  \ss\tt\, ij}$ ($i,j=1,2$)
given by
$$\bar
R_{ ij}= \pmatrix{
  R  &  *R   \cr
 R*&  *R* \cr
}
\eqn\rsq$$
satisfying
$$\bar
R_{\mm\nn  \ss\tt\, ij}=\bar
R_{  \ss\tt\mm\nn\, ji}
\eqn\abc$$
together with the constraint
$$
\bar R_{\mm\nn  \ss\tt\, ij}= J_{i}{}^k(*\bar R)_{\mm\nn  \ss\tt\, kj}
\eqn\rcon$$
with $J$ given by \jiss.

This can be generalised to 
consider tensors $\bar
R_{\mm\nn  \ss\tt\, ij}$ satisfying \rcon\ and
allow $J$ to be any matrix satisfying 
$J^{2}=-1$, with different choices determining different interactions, 
so that \rsq\ is generalised. (The choice \jiss\ recovers \rsq.)
Then if the curvatures satisfy
the first Bianchi identities
$$\bar
R_{[\mm\nn  \ss]\tt\, ij}=
0\eqn\brbo$$
the constraint \rcon\ implies 
the field equations
$$\eta^{\nn\tt}\bar
R_{\mm\nn  \ss\tt\, ij}=0\eqn\brfo$$
and 
$$\bar
R_{\mm\nn  \ss\tt\, ij}=\bar
R_{  \ss\tt\mm\nn\, ij}
\eqn\abc$$
and $$\bar
R_{\mm\nn  \ss\tt\, ij}=\bar
R_{  \mm\nn\ss\tt\, ji}
\eqn\abc$$
so that
$R_{ij}$ is a symmetric $2\times 2 $ matrix.

A particularly interesting choice of $J$ is that given by \jis, so that 
one is introducing a gravitational \lq theta-angle' as well as the 
usual
gravitational coupling constant.
This is the system that arises from the dimensional reduction of the 
free (4,0) conformal multiplet in 5+1 dimensions [\csto].
Then \rcon\ has an 
$SL(2)$ symmetry
under which 
$\bar R_{\mm\nn  \ss\tt\, ij}$ transforms as a 2nd rank  tensor 
$$\bar R_{\mm\nn  \ss\tt } \to S\bar R_{\mm\nn  \ss\tt }S^{t}\eqn\symten$$
(which is a symmetric tensor if \brbo\ holds)
while $J$ transforms as \jtran.
This $SL(2)$ is a gravitational analogue of the S-duality of the 
electromagnetic field and mixes the field equations \brfo\ with the 
Bianchi identities \brbo.
Moreover, \rcon\ then implies the $SL(2)$-invariant constraint 
$$
\gg^{ij}\bar R_{\mm\nn  \ss\tt\, ij}= 0
\eqn\rtr$$
so that $R_{ij}$ is a symmetric matrix with vanishing trace 
\rtr.

Writing the components of $R_{ij}$ as in [\csto] as
$$
 \bar
R_{\mm\nn  \ss\tt\, ij}= \pmatrix{
  R_{\mm\nn  \ss\tt } & \ti R_{\mm\nn  \ss\tt }  \cr
\ti R_{\mm\nn  \ss\tt }& \hat R_{\mm\nn  \ss\tt } \cr
}
= \pmatrix{
  R_{\mm\nn  \ss\tt } & S_{ \ss\tt \mm\nn  }  \cr
S_{\mm\nn  \ss\tt }& G_{\mm\nn  \ss\tt } \cr
}
\eqn\abc$$
with
 $J$   given by \jis,
the Bianchi identities imply that these are the curvature tensors for 
the gravitons
$$
\bar h_{\mm\nn\, ij}= \pmatrix{
h_{\mm\nn}  & \ti h_{\mm\nn}  \cr
\ti h_{\mm\nn}& \hat h_{\mm\nn} \cr
}= \pmatrix{
h_{\mm\nn}  & D_{\mm\nn}  \cr
D_{\mm\nn}& C_{\mm\nn} \cr
}
\eqn\abc$$
Then \rtr\ implies
$$ \hat R_{mnpq } -2 \tt_1 \ti R_{mnpq }+ |\tt|^2 R_{mnpq }=0
\eqn\rrtr$$
so that
$$ \hat h _{mn} -2 \tt_1 \ti h _{mn}+ |\tt|^2 h _{mn}=0
\eqn\abc$$
up to gauge transformations.

 The duality constraint \rcon\
gives the following relations between the curvature tensors (suppressing
the spacetime indices)
$$\eqalign 
{R&={1\over \tt_2}( -*\ti R +\tt_1*  R)
\cr
\ti R&={1\over \tt_2}( -* \hat R +\tt_1 * \ti  R)
\cr
\hat
R&={1\over \tt_2}( |\tt|^2*\ti R -\tt_1  *\hat R)
\cr}
\eqn\rrcon$$
This implies that $\ti R,\hat R$ are given in terms of $R$ by
$$\eqalign{\ti R&= {1\over g^2}*R - \th R
\cr 
\hat R&= 2 \tt_{1}\tt_{2}*R - (\tt_{1}^{2}-\tt _{2}^{2}) R
\cr}
 \eqn\risas$$
 so that only one of the three gravitons $h,\ti h, \hat h$ is 
 independent.
 The one remaining independent graviton can be taken to be $h$ with 
 action
 $$S={1\over 2l^{2}}\int d^{4}x   \, 
(\sqrt g R)_{quad}
\eqn\abc$$
where 
the Planck length is given by
$$l=\sqrt{V} g^{2}
\eqn\abc$$
and $(\sqrt g R)_{quad}$ is the Einstein action truncated to terms
quardatic in
$h_{mn}$. There are two dimensionless 
coupling constants, $g,\th$. While $g$ can be  absorbed into the 
the gravitational coupling $l$, there is the interesting possibility 
of introducing a gravitational $\th$-parameter.

The duality transformations \symten\ take $R$ to linear combinations of 
$R,\ti R,\hat R$ which through \risas\ can be written as linear 
combinations of $R,*R$.
The action of the $SL(2,\Z)$ element
$$S=\pmatrix{
0 & 1 \cr
-1 & 0\cr
}\eqn\abc$$
is to take 
$$\tt \to - {1\over \tt}
\eqn\abc$$
and 
$$ F \to \ti F, \qq \ti F_2 \to - F
\eqn\abc$$
while 
$$ R \to \hat R, \qq \hat R \to R , \qq \ti R \to - \ti R
\eqn\abc$$
Note that this preserves the constraint \rrtr.
For self-dual $F_{i}$ satisfying \fsd, the transformation is the standard duality 
transformation
$$F \to {1\over g^2}*F  + \th F 
\eqn\abc$$
while for self-dual $\bar R$ satisfying \rrcon\
$$R_{mnpq} \to
2 \tt_{1}\tt_{2}(*R)_{mnpq} - (\tt_{1}^{2}-\tt _{2}^{2}) R_{mnpq}
\eqn\abc$$
For $\th=0$, this takes $g \to 1/g$ and so relates strong and weak 
coupling regimes. Both the linear gravity and Maxwell theory are 
self-dual, with the strong coupling regime described by the same 
theory.

\refout

\bye